\documentclass[twocolumn,english,aps,prl,showpacs]{revtex4}
\usepackage[T1]{fontenc}
\usepackage[latin1]{inputenc}
\usepackage{amsmath}
\usepackage{graphicx}
\usepackage{amssymb}

\makeatletter
\@ifundefined{textcolor}{}
{%
 \definecolor{BLACK}{gray}{0}
 \definecolor{WHITE}{gray}{1}
 \definecolor{RED}{rgb}{1,0,0}
 \definecolor{GREEN}{rgb}{0,1,0}
 \definecolor{BLUE}{rgb}{0,0,1}
 \definecolor{CYAN}{cmyk}{1,0,0,0}
 \definecolor{MAGENTA}{cmyk}{0,1,0,0}
 \definecolor{YELLOW}{cmyk}{0,0,1,0}
 }

\makeatother

\makeatother

\usepackage{babel}

\begin{document}

\title{Virial expansion for a strongly correlated Fermi gas}

\author{Xia-Ji Liu$^{1,2}$, Hui Hu$^{1,2,3}$ and Peter D. Drummond$^{1}$}

\affiliation{$^{1}$\ ARC Centre of Excellence for Quantum-Atom Optics, Centre
for Atom Optics and \\
 Ultrafast Spectroscopy, Swinburne University of Technology, Melbourne
3122, Australia, \\
 $^{2}$\ Department of Physics, University of Queensland, Brisbane,
Queensland 4072, Australia, \\
 $^{3}$\ Department of Physics, Renmin University of China, Beijing
100872, China}

\date{\today{}}
\begin{abstract}
Using a high temperature virial expansion, we present a controllable
study of the thermodynamics of strongly correlated Fermi gases near
the BEC-BCS crossover region. We propose a practical way to determine
the expansion coefficients for both harmonically trapped and homogeneous
cases, and calculate the third order coefficient $b_{3}(T)$ at finite
temperatures $T$. At resonance, a $T$-independent coefficient $b_{3,\infty}^{\hom}\approx-0.29095295$
is determined in free space. These results are compared with a recent
thermodynamic measurement of $^{6}$Li atoms, at temperatures below
the degeneracy temperature, and with Monte Carlo simulations. 
\end{abstract}


\maketitle
Strongly correlated Fermi gases are of wide interest and underlie many
unanswered problems in quantum many-body systems, ranging from neutron
stars, hadrons and quark matter through to high $T_{c}$ superconductors
\cite{review}. Recent investigations of Feshbach resonances in ultracold
atomic Fermi gases have opened new, quantitative opportunities to
address these challenges\cite{review}. A great deal of theoretical
work has been carried out for this simple, well-controlled case of
a strongly interacting yet low density Fermi gas, which is known as
the unitarity limit. However, a profound understanding is plagued
by the large interaction strength, for which the use of perturbation
theory requires infinite order expansions. Numerically exact quantum
Monte Carlo simulations are also less helpful than one might expect
\cite{akkineni,bulgac,burovski}. Due to the Fermi sign problem\cite{akkineni},
computer simulations are often restricted to small samples, and are
therefore difficult to extrapolate to the thermodynamic limit.

In this Letter, we approach this problem by using a \emph{controllable}
virial expansion study of \emph{trapped} strongly interacting Fermi
gases at high temperatures. We focus on the low-density physics which
is described by an effective S-wave contact potential. Our expansion
has a small parameter. The fugacity \[
z=\exp(\mu/k_{B}T)\ll1\]
 is small because the chemical potential $\mu$ diverges logarithmically
to $-\infty$ at large temperatures $T$. The virial expansion up
to the second virial coefficient was applied by Ho and Mueller to
explore the universal thermodynamics of a homogeneous Fermi gas at
unitarity \cite{ho2004b}. Here we extend this to the third order
coefficient. Most importantly, we present a practical theoretical
strategy which can even be extended beyond third order. Surprisingly,
we find that the simplest theoretical route to calculating these higher
order coefficients is via the use of exact solutions for the energy
eigenstates of harmonically trapped clusters. This gives a unified
approach to calculating virial coefficients in both trapped and untrapped
cases.

In reality, the strongly interacting spin-1/2 $^{6}$Li and $^{40}$K
fermionic gases were realized by tuning a magnetic field across a
resonance \cite{review}. Extensive experiments have studied the crossover
from the BCS limit (Cooper pairing of atoms) to the BEC limit (Bose-Einstein
condensation of diatomic dimers). The most interesting region lies
at the middle of crossover, where the two-body S-wave scattering
length $a$ becomes much larger in magnitude than the inverse Fermi
vector $1/k_{F}$ \cite{ohara}. Fascinating phenomena may occur in
this `unitarity limit' \cite{ho2004a,thomas}, such as the observed
scale-invariant, universal thermodynamic behaviour \cite{stewart,luo,natphys}.
Accurate, high-order virial coefficients provide an extremely useful
tool in analysing these experimental results above the superfluid
transition. At the same time, the methods given here may have general
applicability to other strongly interacting systems.

%
\begin{figure}
\begin{centering}
\includegraphics[clip,width=0.45\textwidth]{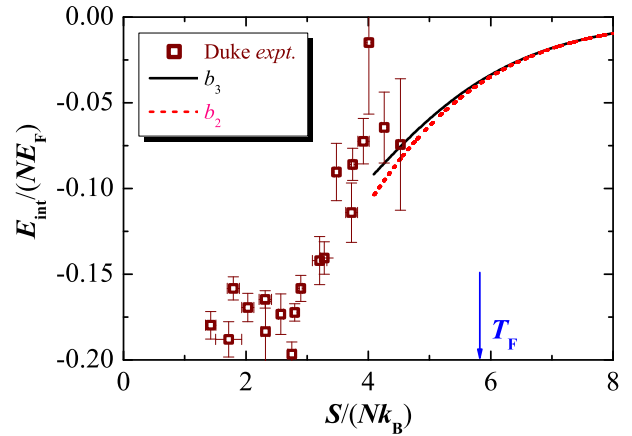} 
\par\end{centering}

\caption{(color online). Predicted interaction energy as a function of entropy
at unitarity, compared with experimental data from Duke University
\cite{luo}. The solid (dashed) line shows the contribution up to
the third (second) order virial coefficient. The arrow indicates
the degeneracy point.}

\label{fig1} 
\end{figure}


In what follows, we first introduce a practical way to calculate the
$n$-th virial coefficients $b_{n}(T)$ and, by solving exactly the
two- and three-particle problems, determine the second and third virial
coefficients in an isotropic harmonic trap at the BCS-BEC crossover.
We then focus on the unitarity limit and calculate the energy and
entropy of a trapped gas using the virial expansion method.

Our main result is summarized in Fig. 1, which shows the comparison
of the virial expansion prediction to a recent measurement of the
entropy-dependence of the interaction energy $E_{int}(S)$ at unitarity.
The experiment was carried out for atomic gases of $^{6}$Li atoms
at a broad Feshbach resonance \cite{luo}. We find an excellent agreement
at temperatures \emph{below} the Fermi degeneracy temperature $T_{F}$.
This remarkable result is opposite to the consensus that the virial
expansion is valid at the classical Boltzmann regime with $T\gg T_{F}$.
We suggest that it can be understood by the significant suppression
of higher-order virial coefficients in a harmonic trap.

Even in the absence of a harmonic trap, our method can still be used
to calculate the virial coefficients. We determine a universal coefficient
$b_{3,\infty}^{\hom}\approx-0.29095295$ for a \emph{homogeneous}
Fermi gas at unitarity, in contrast to a recent calculation that obtained
a result with the opposite sign \cite{rupak}. Our resulting equation
of state is in good agreement with existing Monte Carlo results, and
may provide a useful benchmark for testing future quantum Monte Carlo
simulations of strongly interacting Fermi systems at high temperatures.

\textit{Virial expansion}. --- Let us consider the thermodynamic potential
$\Omega=-k_{B}T\ln{\cal Z}$, where ${\cal Z}=Tr\exp[-({\cal H}-\mu{\cal N})/k_{B}T]$
is the grand partition function. At high temperatures, we can rewrite
${\cal Z}$ in terms of the partition functions of clusters, i.e.,
$Q_{n}=Tr_{n}[\exp(-{\cal H}_{n}/k_{B}T)]$ with $n$ denoting the
number of particles in the cluster and $Tr_{n}$ denoting the trace
over $n$-particle states of the proper symmetry; thus we find ${\cal Z}=1+zQ_{1}+z^{2}Q_{2}+\cdots$.
The thermodynamic potential can then be written as, \begin{equation}
\Omega=-k_{B}TQ_{1}\left[z+b_{2}z^{2}+\cdots+b_{n}z^{n}+\cdots\right],\label{ve}\end{equation}
 where the virial coefficients are given by, \begin{eqnarray}
b_{2} & = & \left(Q_{2}-Q_{1}^{2}/2\right)/Q_{1},\label{b2}\\
b_{3} & = & \left(Q_{3}-Q_{1}Q_{2}+Q_{1}^{2}/3\right)/Q_{1},\ etc.\label{b3}\end{eqnarray}
 These equations present a general definition of virial expansion
and is applicable to both homogeneous and trapped systems. The determination
of the $n$-th virial coefficient thus requires full solutions up
to the $n$-body problem. It is convenient to focus on the interaction
effects only and consider $\Delta b_{n}\equiv b_{n}-b_{n}^{(1)}$
and $\Delta Q_{n}\equiv Q_{n}-Q_{n}^{(1)}$, where the superscript
{}``$1$'' denotes the non-interacting systems. We shall calculate
$\Delta b_{2}=\Delta Q_{2}/Q_{1}$ and $\Delta b_{3}=\Delta Q_{3}/Q_{1}-\Delta Q_{2}$.

\textit{Second and third virial coefficients in traps}. --- By solving
the few-body problem exactly, we now evaluate the virial coefficients
in a three-dimensional isotropic harmonic potential $V(\mathbf{r})=m\omega r^{2}/2$,
with a trapping frequency $\omega$ and fermion mass $m$. The partition
function $Q_{1}$ is easily obtained from the single-particle spectrum
of the harmonic potential, $E_{nl}=(2n+l+3/2)\hbar\omega$, and the
single-particle wave function, $R_{nl}(r)Y_{l}^{m}(\theta,\varphi)$.
We find that $Q_{1}=2\exp(-3\tilde{\omega}/2)/[1-\exp(-\tilde{\omega})]^{3}$
with a dimensionless frequency $\tilde{\omega}=\hbar\omega/k_{B}T\ll1$.
The prefactor of $2$ in $Q_{1}$ accounts for the two possible spins
of each fermion.

To solve the two- and three-fermion problems, we adopt a short-range
S-wave pseudopotential for interatomic interactions, in accord with
the experimental situation of broad Feshbach resonances. This can
be replaced by the Bethe-Peierls contact conditions on the wave function
$\psi\left(\mathbf{r}_{1},\mathbf{r}_{2},...,\mathbf{r}_{n}\right)$:
when any particles $i$ and $j$ with unlike spins close to each other,
$r_{ij}=\left|\mathbf{r}_{i}-\mathbf{r}_{j}\right|\rightarrow0$,
$r_{ij}\psi$ satisfies, \begin{equation}
\partial\left(r_{ij}\psi\right)/\partial r_{ij}=-\left(r_{ij}\psi\right)/a.\label{BP}\end{equation}
 Otherwise, the wave function $\psi$ obeys the noninteracting Schrödinger
equation, \begin{equation}
\sum_{i=1}^{n}\left[-\frac{\hbar^{2}}{2m}\mathbf{\nabla}_{\mathbf{r}_{i}}^{2}+\frac{1}{2}m\omega r_{i}^{2}\right]\psi=E\psi.\end{equation}
 Unlike Bose gases, no additional many-particle interaction parameters
are required. This is due to the absence of many-particle S-wave bound
states (i.e., Efimov-like states) for these low density Fermi gases,
which has a physical origin in the Pauli exclusion principle.

The Hamiltonian of two fermions with different spins was solved by
Busch \textit{et al}. \cite{busch}. As the center of mass is separable
for a harmonic trap, we may single out the (single-particle) center-of-mass
energy $E_{c.m.}$ and rewrite the total energy as $E=E_{c.m.}+E_{rel}$.
Following Busch \textit{et al}. \cite{busch}, the relative energy
$E_{rel}=(2\nu+3/2)\hbar\omega$ satisfies, $2\Gamma(-\nu)/\Gamma(-\nu-1/2)=d/a$,
where $d=\sqrt{2\hbar/m\omega}$ is the length scale of the trap,
and the (un-normalized) two-body relative wave function is given by,
$\psi_{2b}^{rel}(\mathbf{r=r}_{2}-\mathbf{r}_{1};\nu)=\exp(-r^{2}/2d^{2})\Gamma(-\nu)U(-\nu,3/2,r^{2}/d^{2})$.
Here, the total angular momentum of $\psi_{2b}^{rel}$ is strictly
zero since only these states do not vanish at $r=0$ and thus are
influenced by the pseudopotential. $\Gamma$ and $U$ are the Gamma
function and confluent hypergeometric function, respectively. It is
readily shown that $b_{2}-b_{2}^{(1)}=\Delta Q_{2}/Q_{1}$ is given
by: $b_{2}-b_{2}^{(1)}=(1/2)\sum_{\nu_{n}}[e^{-(2\nu_{n}+3/2)\tilde{\omega}}-e^{-(2\nu_{n}^{(1)}+3/2)\tilde{\omega}}]$,
where the summation over $E_{c.m.}$ cancels $Q_{1}$ in the denominator,
and $\nu_{n}^{(1)}=0,1,...$ is the $n$-th solution of the relative
energy spectra in the non-interacting limit. At unitarity, the two-body
solutions, $\nu_{n,\infty}=n-1/2$, are known exactly \cite{busch},
leading to \begin{equation}
b_{2,\infty}-b_{2,\infty}^{(1)}=\frac{1}{2}\frac{\exp\left(-\tilde{\omega}/2\right)}{\left[1+\exp\left(-\tilde{\omega}\right)\right]}=\frac{1}{4}-\frac{1}{32}\tilde{\omega}^{2}+\cdots.\label{b2unitarity}\end{equation}

The three-fermion problem was studied analytically by Werner and Castin
\cite{werner} at unitarity, and numerically by Kestner and Duan \cite{kestner}
for arbitrary scattering lengths. Although the calculations are more
involved, the exact solution is intuitively understandable. Let us
skip the trivial center-of-mass motion. Assuming a spin state $\downarrow\uparrow\downarrow$
and using the Jacobi coordinates $\mathbf{r}=\mathbf{r}_{2}-\mathbf{r}_{1}$
and $\mathbf{\rho}=(2/\sqrt{3})[\mathbf{r}_{3}-(\mathbf{r}_{1}/2+\mathbf{r}_{2}/2)]$
as shown in the inset of Fig. 2, the three-body relative wave function
$\psi_{3b}^{rel}(\mathbf{r},\mathbf{\rho})$ can be written as \begin{equation}
\psi_{3b}^{rel}=\left(1-{\cal P}_{13}\right)\sum\nolimits _{n}a_{n}R_{nl}\left(\rho\right)Y_{l}^{m}\left(\hat{\rho}\right)\psi_{2b}^{rel}(\mathbf{r};\nu_{n}),\end{equation}
 which is simply the summation of products of the eigenstate of the
paired particles 1 and 2, $\psi_{2b}^{rel}(\mathbf{r};\nu_{n})$,
and of the eigenstate of particle 3 relative to the pair, $R_{nl}\left(\rho\right)Y_{l}^{m}\left(\hat{\rho}\right)$.
The value of $\nu_{n}$ for each index {}``$n$'' is uniquely determined
from energy conservation: $E_{rel}/\hbar\omega=(2n+l+3/2)+(2\nu_{n}+3/2)$
and should not be confused with the solutions for the two-body relative
energy. The relative wave function $\psi_{3b}^{rel}$ has a \emph{well-defined}
total relative angular momentum of the 3 particles with quantum numbers
$l$ and $m$. The operator ${\cal P}_{13}$ ensures the correct exchange
symmetry of the wave function. This introduces correlations between
the $1-2$ pair and the remaining particle 3 and thus a hybridization
as parametrized by $a_{n}$. We solve the eigenstate {}``$a_{n}$''
and eigenvalue $E_{rel}$ by imposing the Bethe-Peierls boundary condition
Eq. (\ref{BP}). We find, \begin{equation}
\frac{2\Gamma(-\nu_{n})}{\Gamma(-\nu_{n}-1/2)}a_{n}+C_{nm}a_{m}=\left(\frac{d}{a}\right)a_{n},\label{eigen3b}\end{equation}
 where the (symmetric) matrix $C_{nm}\equiv[(-1)^{l}/\sqrt{\pi}]\times\int_{0}^{\infty}d\rho\rho^{2}R_{nl}\left(\rho\right)R_{ml}\left(\rho/2\right)\psi_{2b}^{rel}(\sqrt{3}\mathbf{\rho}/2;\nu_{m})$
arises from the exchange operator ${\cal P}_{13}$. Without $C_{nm}$
we have a three-body problem of un-correlated pair and single particle.
We label the relative energy in this case as $\bar{E}_{rel}$ and
calculate it directly from the two-body relative energy.

%
\begin{figure}
\begin{centering}
\includegraphics[clip,width=0.45\textwidth]{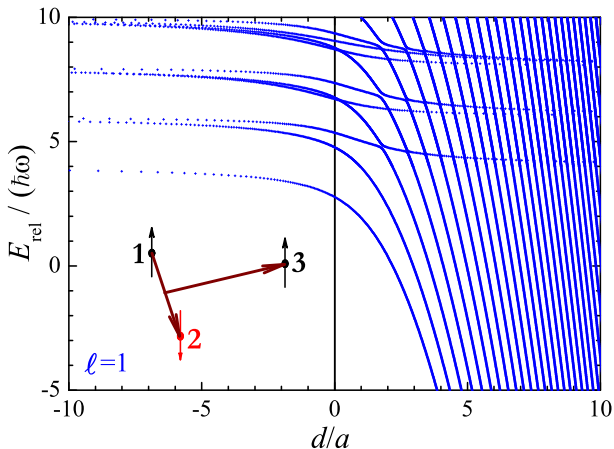} 
\par\end{centering}

\caption{(color online). Relative energy levels of a three-fermion system at
the ground state section ($l=1$).}

\label{fig2} 
\end{figure}


We have solved Eq. (\ref{eigen3b}) numerically for $10^{4}$ energy
levels $E_{rel}$ at different relative angular momenta $l$ and have
checked that at unitarity our results agree exactly with the analytic
spectrum in Ref. \cite{werner}, with relative numerical errors typically
$<10^{-6}$ . Fig. 2 shows how the relative energy spectrum evolves
from the BCS to the BEC side in the subspace of $l=1$.

To calculate the third virial coefficient using $b_{3}-b_{3}^{(1)}=\Delta Q_{3}/Q_{1}-\Delta Q_{2}$,
we notice that the spin states of $\downarrow\uparrow\downarrow$
and $\uparrow\downarrow\uparrow$ contribute equally to $Q_{3}$.
Also, $Q_{1}$ in the denominator is canceled by the summation over
$E_{c.m.}$, and the term $-\Delta Q_{2}$ is cancelled by the difference
between $\bar{E}_{rel}$ and the noninteracting energy $E_{rel}^{(1)}$.
Thus, the third virial coefficient is determined solely by the exchange
correlation, so that $b_{3}-b_{3}^{(1)}=\sum\exp\left(-E_{rel}/k_{B}T\right)-\sum\exp\left(-\bar{E}_{rel}/k_{B}T\right)$,
where the summation is performed over all possible three-body states
that are affected by interactions. At unitarity, we obtain: \begin{equation}
b_{3,\infty}-b_{3,\infty}^{\left(1\right)}=-0.06833960+0.038867\tilde{\omega}^{2}+\cdots.\label{b3unitarity}\end{equation}

%
\begin{figure}
\begin{centering}
\includegraphics[clip,width=0.45\textwidth]{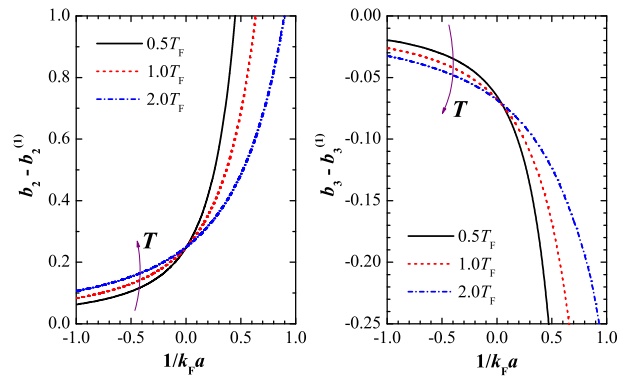} 
\par\end{centering}

\caption{(color online). The second and third virial coefficients as a function
of the interaction parameter $1/k_{F}a$. We have used a total number
of atoms $N=100$, leading to $\tilde{\omega}=(3N)^{-1/3}\approx0.15$
at $T=T_{F}$.}

\label{fig3} 
\end{figure}


The second and third virial coefficients through the crossover is
given in Fig. 3 at three typical temperatures. Here we consider a
gas with $N=100$ atoms and scale the inverse scattering length using
the Fermi vector at the trap center, $k_{F}=(24N)^{1/6}/(d/\sqrt{2})$.
The temperature is given in units of Fermi temperature $T_{F}=E_{F}/k_{B}=(3N)^{1/3}(\hbar\omega/k_{B})$.
All the curves with distinct temperatures cross at $a\rightarrow\pm\infty$.
This is the manifestation of universal behavior anticipated if there
is no any intrinsic length scale. However, the characteristic length
scale $d$ of harmonic traps brings a small (non-universal) temperature
dependence that decreases as $N^{-2/3}$, shown by the terms $\tilde{\omega}^{2}$
in Eqs. (\ref{b2unitarity}) and (\ref{b3unitarity}).

\textit{High-T thermodynamics in traps}. --- We are ready to investigate
the thermodynamics of a strongly interacting Fermi gas at high temperatures.
At unitarity, the energy $E$ and entropy $S$ in the limit of small
$\tilde{\omega}$ or large $N$ can be calculated according to the
universal relations \cite{ho2004a,thomas}, \begin{eqnarray}
E & = & -3\alpha\Omega/2,\label{etrap}\\
S & = & -\left(3\alpha/2+1\right)\Omega/T-k_{B}N\ln z,\label{strap}\end{eqnarray}
 together with Eq. (\ref{ve}) for $\Omega$ and the number identity
$N=-\left(\partial\Omega/\partial\mu\right)=Q_{1}\left[z+2b_{2}z^{2}+\cdots\right]$,
and the virial coefficients shown in Eqs. (\ref{b2unitarity}) and
(\ref{b3unitarity}). Here, $\alpha=2$ for a harmonically trapped
gas. We also calculate the equation of state $E_{IG}(S)$ of an ideal
non-interacting Fermi gas, using $b_{n}^{\left(1\right)}=\left(-1\right)^{n+1}[1/n^{4}-\tilde{\omega}^{2}/(8n^{2})+\cdots]$.

Fig. 1 shows the predicted interaction energy $E_{int}=E-E_{IG}$
as a function of entropy, compared to the experimental data reported
by Luo \textit{et al.} \cite{luo}. We find a rapid convergence of
expansion, even below the degeneracy temperature $T_{F}$, with excellent
agreement between theory and experiment: the virial expansion is applicable
to a trapped Fermi gas even at $T<T_{F}$.

This remarkable observation is counter-intuitive, as the virial expansion
is generally believed to be useful at the Boltzmann regime with $T\gg T_{F}$.
This occurs because there is a significant reduction of higher-order
virial coefficients in harmonic traps. Consider the thermodynamic
potential of a harmonically trapped gas in the local density approximation,
$\Omega=\int\Omega\left(\mathbf{r}\right)d\mathbf{r}\propto\int d\mathbf{r}[z\left(\mathbf{r}\right)+b_{2,\infty}^{\hom}z^{2}(\mathbf{r})+\cdots+b_{n,\infty}^{\hom}z^{n}(\mathbf{r})+\cdots]$,
where $z(\mathbf{r})=z\exp[-V\left(\mathbf{r}\right)/k_{B}T]$ is
a local fugacity with the local chemical potential $\mu(\mathbf{r})=\mu-V\left(\mathbf{r}\right)$.
It is readily seen on spatial integration that the universal ($T$-independent)
part of the trapped virial coefficient is, \begin{equation}
b_{n,\infty}(\text{universal})=\left(\frac{1}{n^{3/2}}\right)b_{n,\infty}^{\hom}.\label{trap_vs_hom}\end{equation}
 Therefore, the higher density of states in traps suppresses the higher
order virial coefficients, leading to an improved convergence of the
expansion at low temperatures.

%
\begin{figure}[htp]

\begin{centering}
\includegraphics[clip,width=0.45\textwidth]{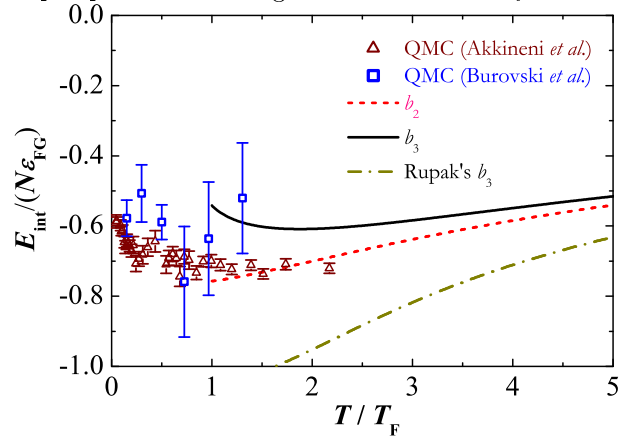} 
\par\end{centering}

\caption{(color online). Temperature dependence of interaction energy of a
homogeneous gas at unitarity, obtained using different virial coefficients.
For comparison, two quantum Monte Carlo data reported in Refs. \cite{akkineni}
and \cite{burovski} are shown. A finite range of interactions has
been used in Ref. \cite{akkineni}, which may lead to a systematic
downshift in energies. We list also the result calculated by using
Rupak's $b_{3,\infty}^{\hom}\approx1.11$. }

\label{fig4} 
\end{figure}


\textit{High-T thermodynamics in free space}. --- Using relation (\ref{trap_vs_hom}),
we may determine the third virial coefficient in free space: $b_{3,\infty}^{\hom}\approx-0.29095295$.
This does not agree with a previous field-theoretic calculation reported
by Rupak \cite{rupak}, $b_{3,\infty}^{\hom}\approx1.11$. As well,
we may calculate the high-$T$ thermodynamics in free space, by taking
$\alpha=1$ and $Q_{1}=2V(mk_{B}T/2\pi\hbar^{2})^{3/2}$ in Eqs. (\ref{etrap})
and (\ref{strap}). Fig. 4 presents the interaction energy $E_{int}$
of a homogeneous Fermi gas at unitarity as a function of temperature.
For comparison, we also show the results of two quantum Monte-Carlo
simulations. The virial expansion in free space seems to converge
at $T>2T_{F}$.

In conclusion, we have shown that the virial expansion converges rapidly
for a degenerate, resonant Fermi gas in a harmonic trap. This allows
us to investigate the thermodynamics in a controllable way. We have
proposed a practical method to obtain the third virial coefficient
throughout BCS-BEC crossover. Higher order coefficients are calculable
in a similar manner, and may hold the prospect of revealing the exact
thermodynamics of resonant Fermi gases in the deep degenerate regime.
The current work provides a useful benchmark on testing future experiments
and quantum Monte-Carlo simulations on strongly interacting Fermi
gases.

This work was supported in part by ARC Center of Excellence, NSFC
Grant No. 10774190, and NFRPC Grant Nos. 2006CB921404 and 2006CB921306.

\end{document}